\documentclass[aps,pre,reprint,showpacs,superscriptaddress]{revtex4-1}
\usepackage{epsfig}
\usepackage{natbib}
\usepackage{ulem}
\usepackage{xspace}
\usepackage{xcolor}
\usepackage{bbold}
\usepackage{appendix}

\begin{document}

\title{ Regularized Boltzmann-Gibbs statistics for a Brownian particle \\ in a non-confining field  }

\author{Lucianno  Defaveri}
\affiliation{Department of Physics, PUC-Rio, Rio de Janeiro, Brazil}
\author{Celia Anteneodo}
\affiliation{Department of Physics, PUC-Rio, Rio de Janeiro, Brazil}
\affiliation{Institute of Science and Technology for Complex Systems, Rio de Janeiro, Brazil.}
\author{David A. Kessler}
\affiliation{Department of Physics, Institute of Nanotechnology and Advanced Materials, Bar-Ilan University, Ramat-Gan
52900, Israel}
\author{Eli Barkai}
\affiliation{Department of Physics, Institute of Nanotechnology and Advanced Materials, Bar-Ilan University, Ramat-Gan
52900, Israel}

\begin{abstract}
We consider an overdamped Brownian particle subject to an asymptotically 
flat potential with a trap of depth $U_0$  around the origin. 
When the temperature is small compared to the trap depth ($\xi=k_B T/U_0 \ll 1$), 
there exists a range of timescales over which physical observables remain 
practically constant. This range can be very long, 
of the order of the Arrhenius factor ${\rm e}^{1/\xi}$.   
For these quasi-equilibrium states, the usual Boltzmann-Gibbs recipe does not work, 
since the partition function is divergent due to the flatness of the potential at long distances. 
However, we show that the standard Boltzmann-Gibbs (BG) statistical framework and thermodynamic relations 
can still be applied through  proper regularization. 
This can be  a valuable tool for the analysis of   metastability in the non-confining potential 
fields that characterize a vast number of systems. 
 
\end{abstract}
\maketitle
\section{Introduction}

In nature, a system coupled to a thermal environment may not be confined indefinitely. A well known example is the hydrogen atom coupled to a thermal bath
at temperature $T$. As noted by Fermi~\cite{Fermi}, the partition function of the hydrogen atom
diverges~\cite{Plastino}.  This is due to the fact that the Coulomb potential is asymptotically
flat, and hence non-binding in a thermal setting. 
Similarly, the barometric formula is an excellent practical approximation  to describe the density of 
particles in the atmosphere, at least in the vicinity of earth~\cite{book}.  
However, as in the Coulomb case, the gravitational field of the earth 
is not sufficient to maintain an atmosphere in equilibrium for an infinite time.
Another well-known example is when a Kramers reaction coordinate is  in the vicinity of a metastable
state~\cite{Kramers,Redner,Chupeau}, 
where the system can stay for a very long time, 
although it is eventually destined to escape. 
In all these examples, the partition function of the single
particle is divergent, and hence we cannot apply the usual toolbox 
of equilibrium BG statistical mechanics even if the system appears
 to be in a thermal steady state~\cite{ErezPRL,ErezCond,Ryabov,Ryabov2,Neri}. 
On the other hand, as we shall see, if the potential has a deep minimum when compared to the temperature, the system may attain a quasi-equilibrium (QE) state, in the sense that thermodynamic observables, like the energy and the entropy are almost time 
independent.  
Then, the question arises whether we can somehow apply 
concepts of equilibrium thermodynamics to these states, relating the microscopic 
dynamics to thermodynamics, through some kind of partition function.  
In doing so, however, we cannot simply use the
the BG prescription, since  the 
partition function as usually defined is divergent. 
We can still ask if it is possible to regularize the Boltzmann-Gibbs prescription, and how to do it.  

Here our goal is first to show that by regularizing the divergent partition function we are able to calculate the values of  observables in these QE states, and hence provide  a  complete toolbox for non-confining potentials. 
For that aim, we investigate the overdamped dynamics of a single  classical particle coupled to a thermal heat bath with temperature $T$. 
The particle starts at the minimum of the potential field, and then, due to the separation of time scales, one sees that the physical observables  appear stationary for long times. 
The first  theoretical challenge is to define this quasi-equilibrium state precisely. 
This is done with an inflection-point technique, which can be viewed as an extension of extremum principles 
found for ordinary equilibrium.  
Secondly, what are the values of observables in quasi-equilibrium? How can one obtain them given that
 the partition function is infinite? 
One way is to solve the Fokker-Planck equation (FPE); however, this is hard, and we wish to find solutions which are nearly time-independent. 
So, we seek a method that does not consider time explicitly,  invoking principles of ordinary equilibrium 
 statistical  mechanics.  
Finally, we ask: 
 is the structure of thermodynamics still maintained, even in the absence of true equilibrium? 
For this question, we show that a regularized partition function is useful in the evaluation of  free energy, entropy, and energy, 
just like in usual textbooks, but with a new partition function which we call $Z_K$. 

The paper is organized as follows. 
In Sec.~\ref{sec:system}, we describe the kind of systems 
considered, which exhibit QE states as 
defined in Sec.~\ref{sec:QE}. 
In Sec.~\ref{sec:time}, we show how to regularize QE states, when the standard partition function is non-normalizable in the unbounded space. 
An alternative, finite-size approach is presented in Sec.~\ref{sec:bounded} and final comments in Sec.~\ref{sec:final}.

\section{The system \label{sec:system}}

Consider a Brownian particle in one dimension  coupled to a thermal
heat bath. The concepts we will discuss are not modified in higher dimensions. 
We assume that the motion is overdamped and that the Einstein relation
between diffusivity and damping holds. 
The density $P(x,t)$ is then described by the
FPE~\cite{Risken,Gardiner} with the force field $F(x) = - \partial_x V(x)$,  
namely
$\partial_t P(x,t)  = D \bigl( \partial^2_{xx}  -  \partial_x { \frac{ F(x) }{  k_BT }} \bigr) P(x,t)$,
where $D$ is the diffusion coefficient, $T$  the temperature and $k_B$ the Boltzmann constant. 
The  potential field $V(x)$ has a local minimum at $x=0$ and it is assumed to be an even function. 
The key feature is that ${\displaystyle \lim_{x\to \infty}} V(x) = 0$, and that the field is bounded from below. 
A family of potentials that fulfill these conditions, and hence will be used to exemplify the problem, is 
\begin{equation}
V_\mu(x)=-\frac{U_0}{\big( 1+ (x/x_0)^2 \big)^{\mu/2}} \,,
\label{eq:potential}
\end{equation}
with $U_0,\; \mu>0$. 
After suitably scaling variables (namely, $x/x_0\to x$, $Dt/x_0^2\to t$, $ V(x)/U_0\to v(x)$), 
the FPE assumes the non-dimensional form
\begin{equation}
{ \frac{ \partial P(x,t) }{  \partial t }} = 
\frac{\partial^2}{\partial x^2} P(x,t) + \frac{1}{\xi} \frac{\partial}{\partial x} 
\Big( \frac{\partial v(x)}{\partial x} P(x,t) \Big) \,,  
\label{eq:FPE}
\end{equation}
where the only free parameter for a given $\mu$ is the reduced temperature $\xi=k_B T/U_0$.

\section{Quasi-equilibrium}
\label{sec:QE}

\begin{figure}[b!]
\centering
\includegraphics[width=0.5\textwidth]{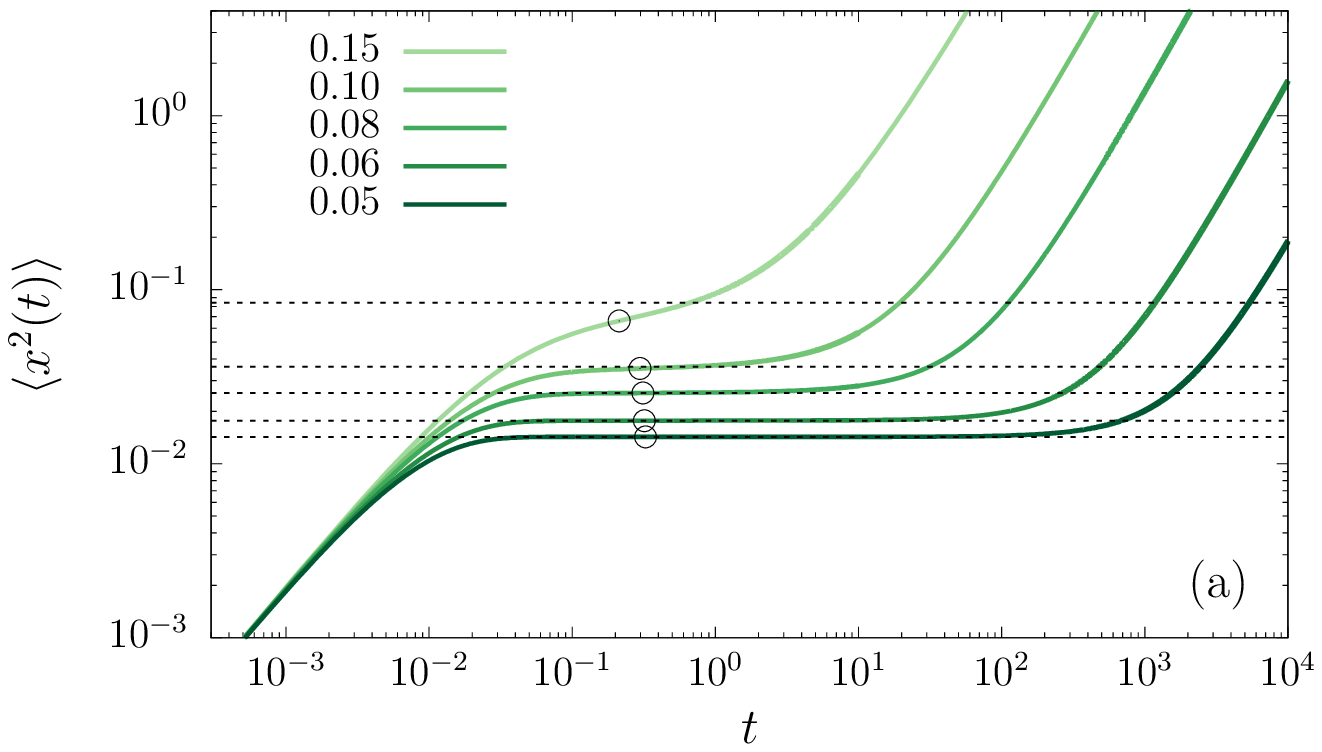}
\includegraphics[width=0.5\textwidth]{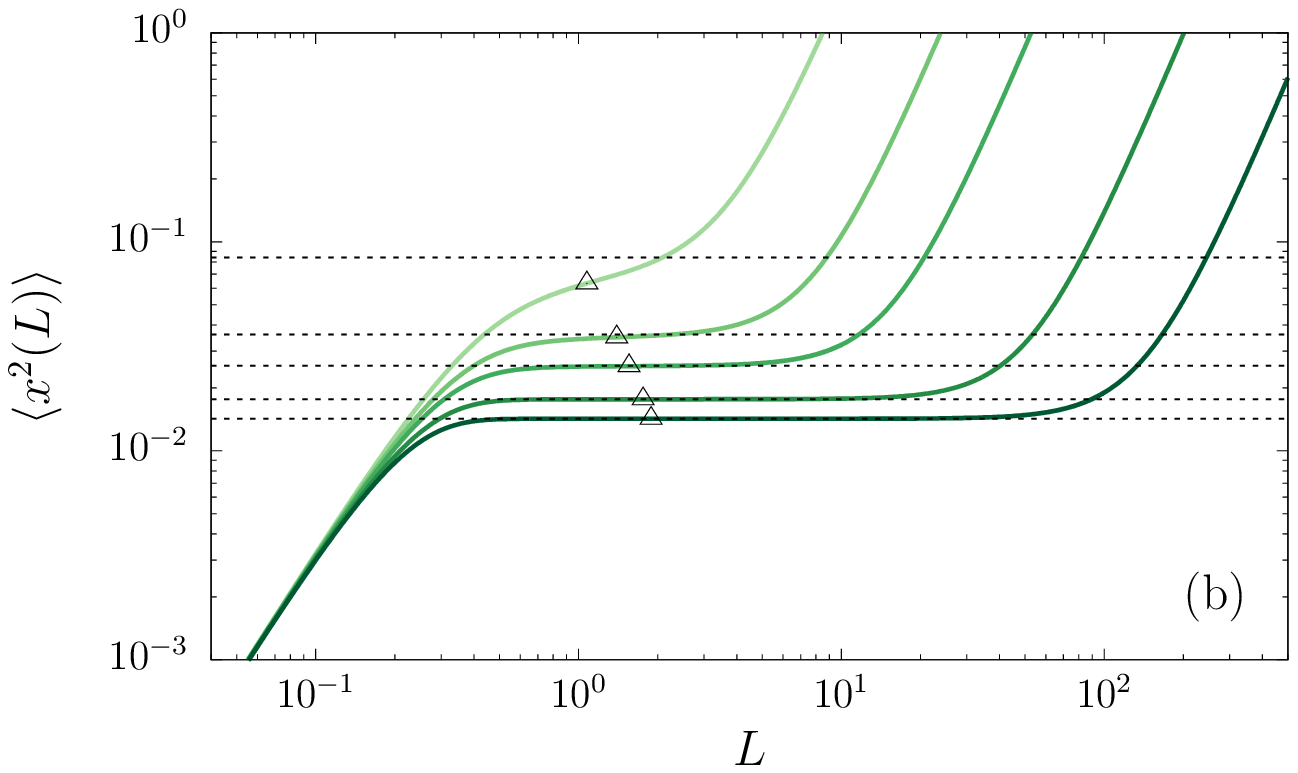}
\caption{  
(a) MSD   $\langle x^2(t) \rangle$ versus time $t$, obtained from the numerical solutions 
of the FPE~(\ref{eq:FPE}), starting from a 
Dirac delta at the origin, with the potential field  $v_4(x)=-1/(1+x^2)^2$, 
for different values of 
the scaled temperature $\xi=k_BT/U_0$ indicated in the legend.
At short and long times, particles move almost in free diffusion, but, 
at intermediate scales, they are in a QE state, 
whose lifetime is longer the smaller is $\xi$. 
We highlight the log-inflection point in the plateau region (open circle),  
defined by Eq.~\ref{eq:loginflection}, which was numerically calculated. 
Dotted horizontal lines (in both panels) are the theoretical prediction 
based on QE Eq.~(\ref{eq:msd}), 
which perfectly match the MSD when stagnation is long lived.
(b) MSD   $\langle x^2(L) \rangle$ versus $L$,  given by Eq.~(\ref{eq:msd_L}), for 
particles subject to  the field  $v_4(x)$ in a finite box $[-L,L]$.
The ($\log L$)-inflection point (open triangle) is 
given by Eq.~(\ref{eq:msd_L-star}). 
}
\label{fig:msd}
\end{figure}

Qualitatively, as seen in Fig.~\ref{fig:msd}a, showing the mean-squared displacement (MSD) 
$\langle x^2(t) \rangle$ vs. time from a direct numerical integration of the FPE, Eq.~(\ref{eq:FPE}), the dynamics of a packet of particles all starting at $x=0$ follows three stages. 
(i) For very short times, the particles spread diffusively and the force is not yet felt. 
Hence here $\langle x^2\rangle \sim 2t$.
(ii) When  the potential well is deep compared to the temperature ($\xi \ll 1$), the system stagnates
 at intermediate time scales and then  $\langle x^2\rangle$ is roughly a constant. 
Importantly, these timescales can be exponentially long, proportional to 
${\rm e}^{1/\xi}$, the Arrhenius  factor.   
(iii) Finally for very  long times, 
the density of particles within the trap will decay to zero as $t^{-1/2}$, 
and the  particles will  diffuse to large distances since 
there are not, in principle, external boundaries to block the diffusion. 

We are interested in the intermediate time scale,
where there is stagnation of the dynamics, 
as seen in the plateau in Fig.~\ref{fig:msd}, which we 
denominate a QE state. In other contexts, QE has alternative definitions~\cite{schwabl,gorban}, 
 or the system is described as being in a long-lived metastable state~\cite{Garrahan}. 
Notice in Fig.~\ref{fig:msd}a that when we increase $\xi$ (by increasing the temperature or reducing the well deepness), 
the QE effect signaled by the stagnation clearly diminishes in strength. 
Similar plots have been reported for systems of particles with Lennard-Jones interactions~\cite{KobAnderson,Berthier2011},  
particles diffusing in porous media~\cite{olivares},
and are also found in single molecule experiments~\cite{superdiff,anomalous}. 
Note however that we do not treat many body effects found in glasses, 
neither anomalous diffusion found in single molecule experiments. 
Moreover, as we will see, similar long-lived quasi-stationary regimes also occur for thermodynamic observables, like the energy and the entropy.

With regard to a more precise definition of quasi-equilibrium,
we note that thermodynamics is formulated via extremum principles, and similarly here
the inflection points of observables can be useful to define a
characteristic QE value, 
as shown for the MSD in Fig.~\ref{fig:msd}a. 
More precisely,  since usually  metastability involves timescales  over many orders of magnitude, and hence it is  analyzed on a logarithmic scale,  we define the $\log(t)$-inflection point through
\begin{equation} \label{eq:loginflection}
    d^2 \langle x^2 \rangle/ d( \log t)^2 =0  \,.
\end{equation}
The values of $\langle x^2 \rangle$, and similarly for other observables,   satisfying this equation at the inflection time is what we use as the QE value of the observable 
(see Fig.~\ref{fig:msd} and further discussion in Sec.~\ref{sec:bounded}). 
This useful  definition will be soon  explored.
The main  idea behind the inflection point
 is that instead of relying on an intuitive  definition of QE, that is, 
when observables  ``nearly'' do not change over time, the inflection point is clearly well defined and lies within the stagnation region.  
Importantly, as shown in Fig.~\ref{fig:msd}, this definition allows us to go from a low temperature phase, where stagnation lasts a few orders of magnitude over time, to higher temperatures where we still have an inflection point, but the stagnation is not so long-lived.
We take the system state at the inflection point as the definition of the QE state. 
Thus, mathematically, while equilibrium is defined through extremization (vanishing of the first derivative) of a thermodynamic potential, 
the vanishing of a second derivative  defines QE. 

Let us remark that, in principle,  one could consider the inflection point  defined through $d^2 \langle x^2 \rangle / d t^2=0$.
This inflection point is always found for the MSD,   
 since $\langle x^2\rangle \sim 2 t$ for short times and similarly for very long times. 
This would imply metastability also for very high temperatures,  which is not natural. The log-inflection point defined above exhibits a critical value as shown in Sec. ~\ref{sec:bounded}. 
In that sense it satisfies the intuitive  demand that QE states are found for finite times and low temperatures.  
To summarize, the inflection  time and the value of the MSD at this inflection point within the stagnation regime can be found numerically or experimentally, and this useful 
quantifier  of QE  is investigated theoretically below.


\section{Time-dependent solution approach}
\label{sec:time}

\begin{figure}[t!]
	\centering
\includegraphics[width=0.49\textwidth]{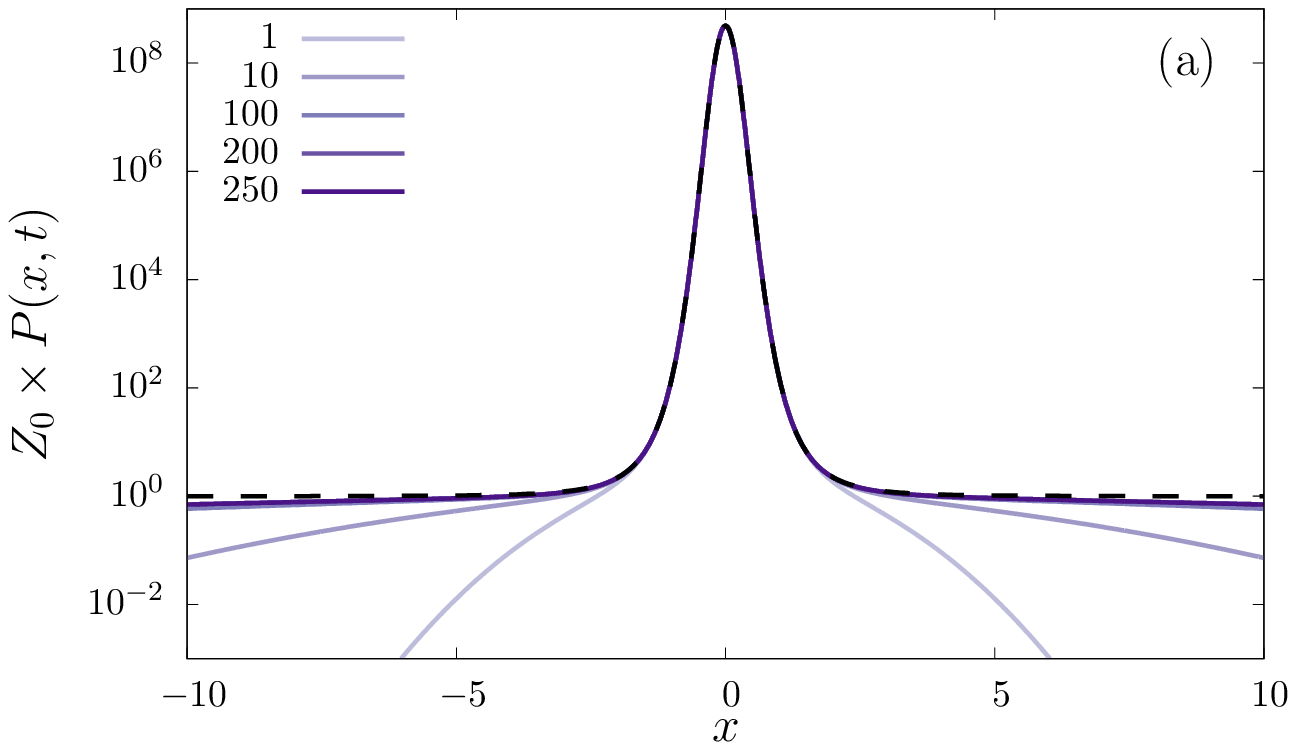}\\
\includegraphics[width=0.49\textwidth]{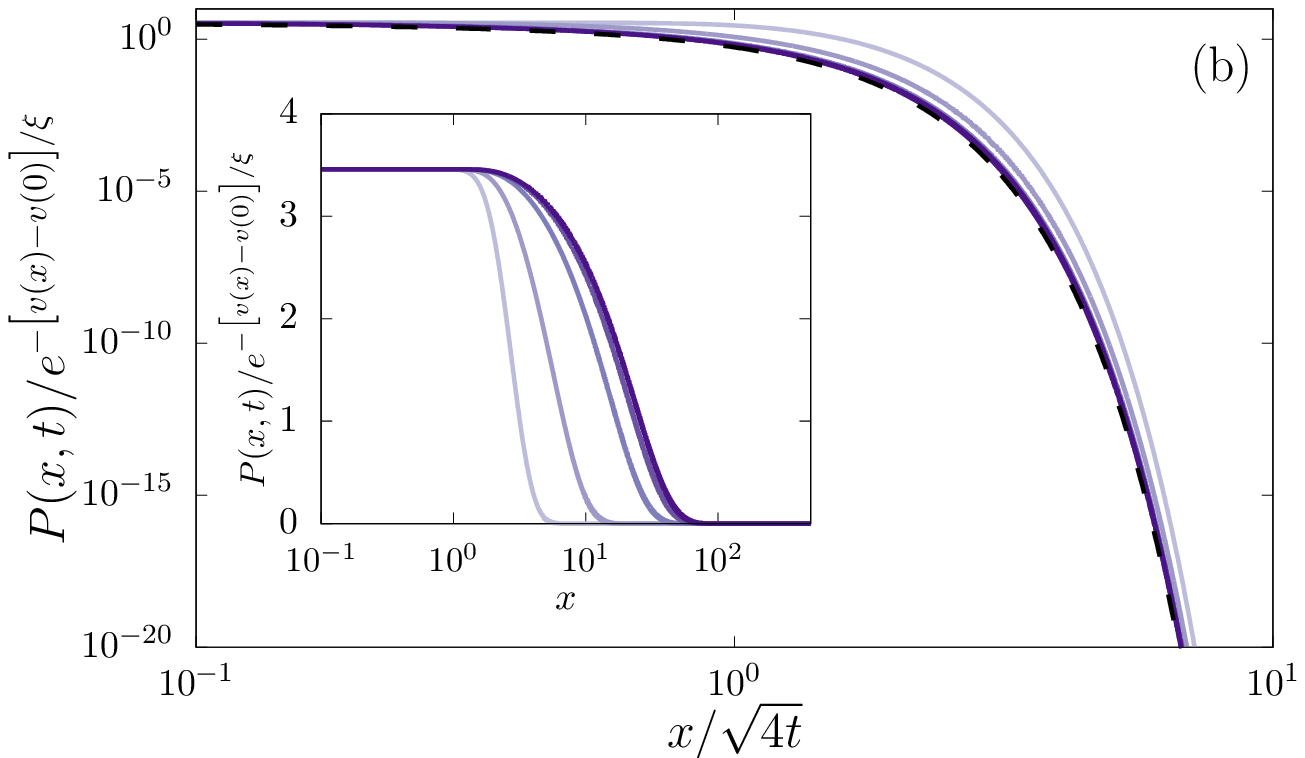}
\caption{ 
Numerical solution of FPE (\ref{eq:FPE}), with potential $v_4(x)$ and $\xi=0.05$, 
vs. $x$ (full lines), after a transient regime starting from a Dirac delta at the origin,  
at different times $t$ indicated in the legend. 
(a)  $Z_0 \times P(x,t)$ vs. $x$ shows that  
the PDF adopts for small $x$ a time-independent shape, 
corresponding to the non-normalizable 
Boltzmann factor  ${\rm e}^{-v(x)/\xi}$, represented by dotted lines, 
while at large $x$ the erfc-like factor imposes a cutoff. 
(b) $P(x,t)/{\rm e}^{-[v(x)-v(0)]/\xi}$ vs. $x/\sqrt{4t}$ 
shows the tail of the density, i.e., the cut-off factor, which tends to the 
${\rm erfc}(x/\sqrt{4t})$ (black dashed line) for long, but not too long, times, 
and the inset highlights that this factor is flat in the region of the effective 
well, all as described by Eq.~(\ref{eq:BC}). 
	}
	\label{fig:pdfs}
\end{figure}

Before presenting an analytical approach to the problem, we show in Fig.~\ref{fig:pdfs} 
the numerical solutions of the FPE (\ref{eq:FPE}), starting from
a Dirac delta at the origin, after a transient. In order
to obtain a QE density $P(x)$, we analyze the FPE at
timescales where a QE is reached, namely, times shorter
than the Arrhenius timescale, but longer than the time it
takes the particle to explore the local minimum.  
Figure~\ref{fig:pdfs}a demonstrates that $P(x,t)$ times a constant $Z_0$ is 
the Boltzmann factor  ${\rm e}^{- v(x)/\xi}$. 
This holds on spatial scales $x<1$ and the theoretical challenge is to find $Z_0$ which is a regularized partition function. 
Clearly, this implies that BG statistics is still a useful concept even though the normalizing partition function diverges. 
Examining the figure in more detail, we see that there are two lengthscales in 
the problem.  The first is the lengthscale of the potential well, which is unity ($x_0$ in the original variable). The second is time-dependent and associated with the
``wings" of the distribution outside the well, which spread out in time.  Since the motion in this region is almost purely 
diffusive, this lengthscale is the diffusion scale $\sqrt{t}$.

We first notice that for the shorter lenghtscale $x \sim {\cal O}(1)$, 
the time-independent QE solution of the FPE  is  $P(x,t) \propto const. \; {\rm e}^{-v (x)/\xi}$.
This is essentially the same as true equilibrium, however 
the constant  must deviate from the inverse standard normalizable partition function  since,
as already mentioned, the latter diverges. 
To determine this constant, we need to find the FPE solution for large distances $ x> 1$, and then match it to that of the inner region where the Boltzmann-like solution is valid. 
Assuming a scaling ansatz for $P(x,t) = t^{\alpha}g(y)$, with $y\equiv x/\sqrt{t}$, the FPE reads to leading order  in $y$
\begin{equation}
\frac{\alpha}{t} g  - \frac{yg'}{2t} = \frac{g''}{t},
\end{equation}
as long as $\mu > 1$, so that the potential terms are higher-order. 
Namely, since we deal here with $x \gg 1$, the force is totally negligible. 
To determine $\alpha$ and $g$, we have to consider the boundary condition for large $y$ and also match this scaling solution to the solution of the FPE for $1 \ll x \ll \sqrt{t}$.   At large $y$, we have to choose the decaying solution, so $g(y) \sim y^{-2\alpha-1} e^{-y^2/4}$. For small $y$ but large $x$, we have to match to the quasi-equilibrium Gibbs solution, which goes to a time-independent constant.  Hence, $\alpha=0$. Thus, in the scaling regime, we have
\begin{equation}
-\frac{yg'}{2} = g'',
\end{equation}
which yields the complementary error function \cite{eigenf}
\begin{equation}
g \sim A \,\textrm{erfc}(y/2).
\end{equation}
Thus, for $x \sim {\cal O}(1)$, $P(x,t) \sim Ce^{-v(x)/\xi}$ and in the diffusive region $P(x,t) \sim A\, \textrm{erfc}(x/\sqrt{4t})$. A uniform approximation which reproduces both behaviors is
\begin{equation}
P(x,t) \simeq C e^{-v(x)/\xi}\, \textrm{erfc}(x/\sqrt{4t}),
\label{eq:BC}
\end{equation}
where we used $\lim_{x\to\infty}V(x)=0$. We see that the erfc gives an effective cutoff at large distances. 
To obtain the factor $C$, we split the normalization condition for the PDF defined 
in Eq.~(\ref{eq:BC}) in terms of an intermediate length scale $\ell$ ($1 \ll \ell \ll \sqrt{t}$), such that  
\begin{eqnarray} \nonumber
 &&\frac{1}{2 C} \simeq
\int_0^\ell {\rm e}^{-v(x)/\xi}dx + \int_\ell^\infty 
{\rm erfc}\Bigl( {x \over   \sqrt{ 4t} } \Bigr)dx 
\\  \nonumber
&\simeq& 
\underbrace{\int_0^\infty ({\rm e}^{-v(x)/\xi}-1)dx}_{\textstyle Z_0/2} - 
\underbrace{\int_\ell^\infty ({\rm e}^{-v(x)/\xi}-1)dx}_{\textstyle R} 
\\ \label{eq:RF}
&& +  \int_0^\infty {\rm erfc} \Bigl( {x  \over \sqrt{ 4t} } \Bigr) dx \,  
\simeq Z_0/2    + {\cal O}(\sqrt{t}) \,, 
\end{eqnarray} 
where we have assumed that the potential decays faster than $1/x$. 
The integral denoted  $Z_0/2$ is the dominant term in Eq.~(\ref{eq:RF}), 
it is time independent and of order ${\rm e}^{1/\xi}$. 
Its integrand is essentially  the Mayer f-function and $Z_0$ is proportional to   
the second virial coefficient from the theory of gases~\cite{Montroll}. 
The  term $R$ scales as $1/\xi$, so that it becomes increasingly negligible 
compared to $Z_0$. 
The ${\rm erfc}$ integral   
grows with time, but here we assume that this diffusive
 length scale is small in the sense that $\sqrt{t} \ll Z_0 $  and so,
 as long as $t$ is not exponentially large, the last term can be neglected  compared to $Z_0$. 
For times  much longer than ${\rm e}^{1/\xi}$,  
the widely discussed infinite ergodic theory applies~\cite{ErezPRL,ErezCond}.
 
To summarize, if the potential field decays faster than $1/x$,  
the PDF  is given by  
$ P(x,t) \simeq {\rm e}^{  - v(x)/\xi} \, 
{\rm erfc}(x/\sqrt{ 4 t})/Z_0$,  for $x>0$.
The cutoff at large $x$ stems from the pure diffusive process arising 
from the vanishing of the force  at large $x$. 
The integral $Z_0$ in the denominator, playing the role of a partition function, is finite.  
For large enough time,  the cutoff factor is unity for distances within the size of the well (see inset of Fig.~\ref{fig:pdfs}b), and we have
\begin{equation}
P_{\rm QE}(x) \sim   \frac{{\rm e}^{  - v(x)/\xi}}{ Z_0 } \, ,
\label{eq:PDF}
\end{equation} 
which resembles the canonical BG law (see Fig.~\ref{fig:pdfs}a). 
Here the regularized normalization  factor $Z_0$ is given 
by Eq.~(\ref{eq:RF}), provided that $\mu>1$.
We will soon show that $Z_0$ is indeed a partition function in the sense that it can be used to find  averaged thermodynamic observables like the energy, 
when the potential decays faster than $1/x$.

When the potential decays slower than $1/x$, namely as $1/x^\mu$, with $0<\mu\le 1$, 
$Z_0$ in Eq.~(\ref{eq:RF}) diverges, and a technical modification of the basic formula is required. 
By adding and subtracting terms in Eq.~(\ref{eq:RF}), we obtain the generalized normalization
\begin{equation} \label{eq:ZR}
Z_K  
 = 2\int_0 ^\infty  \Bigl( {\rm e}^{-v(x)/\xi} - \sigma_K(x;\xi) \Bigr) dx \, ,
\end{equation}
where  
\begin{equation} \label{eq:sigma}
\sigma_K(x;\xi)\equiv \sum_{k=0}^{K}  \bigl(-v(x)/\xi \bigr)^k/k!,
\end{equation} 
with $K=\lfloor 1/\mu \rfloor$ (where $\lfloor\ldots\rfloor$ means floor function), 
ensuring a non-divergent integral.  
In this case Eq.~(\ref{eq:PDF}) is still valid, but 
$Z_0$ is replaced with $Z_K$.

\begin{figure*}[t!]
	\centering
\includegraphics[width=0.47\textwidth]{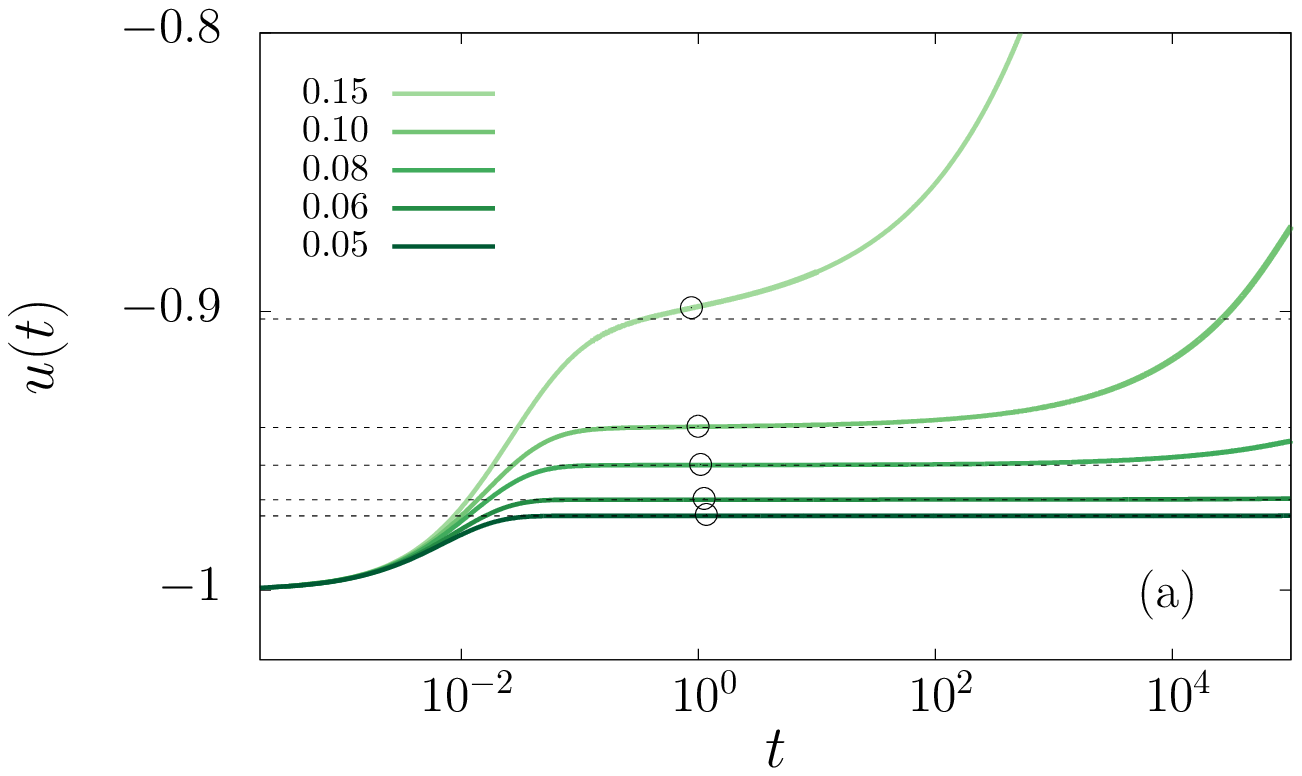}
 \includegraphics[width=0.47\textwidth]{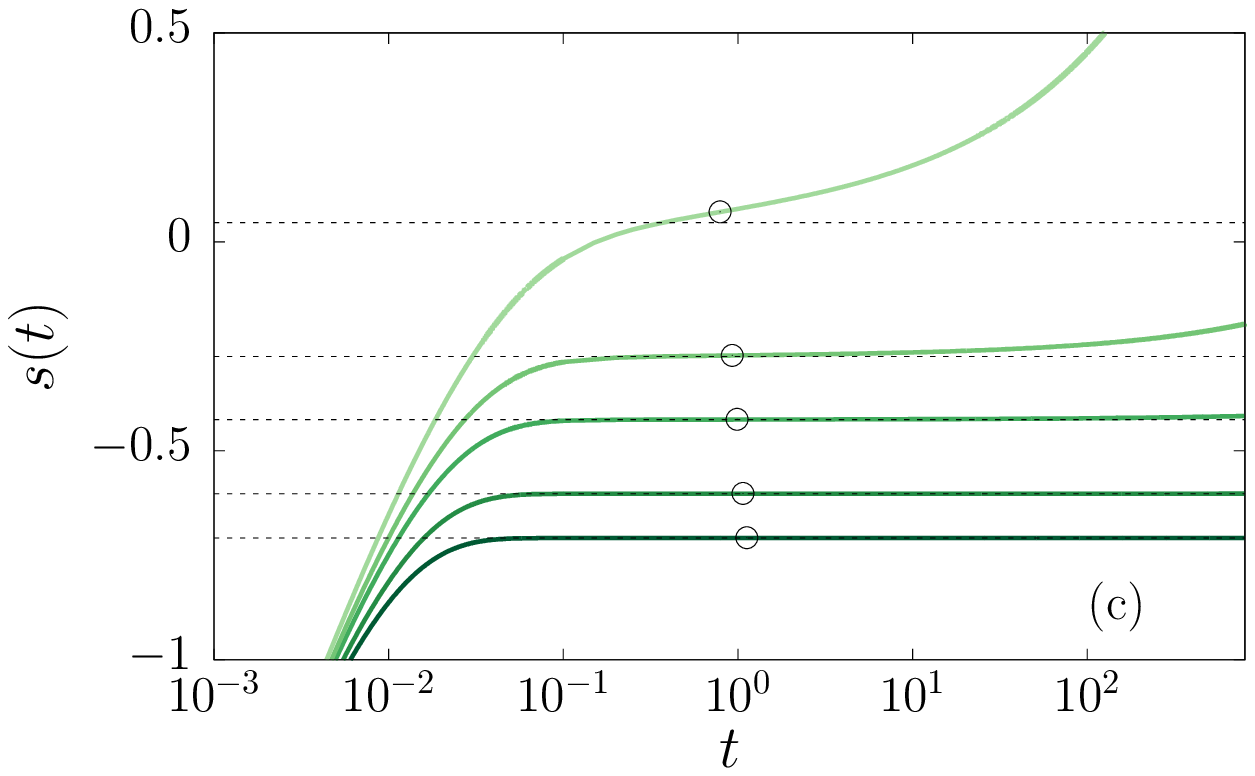}\\
\includegraphics[width=0.47\textwidth]{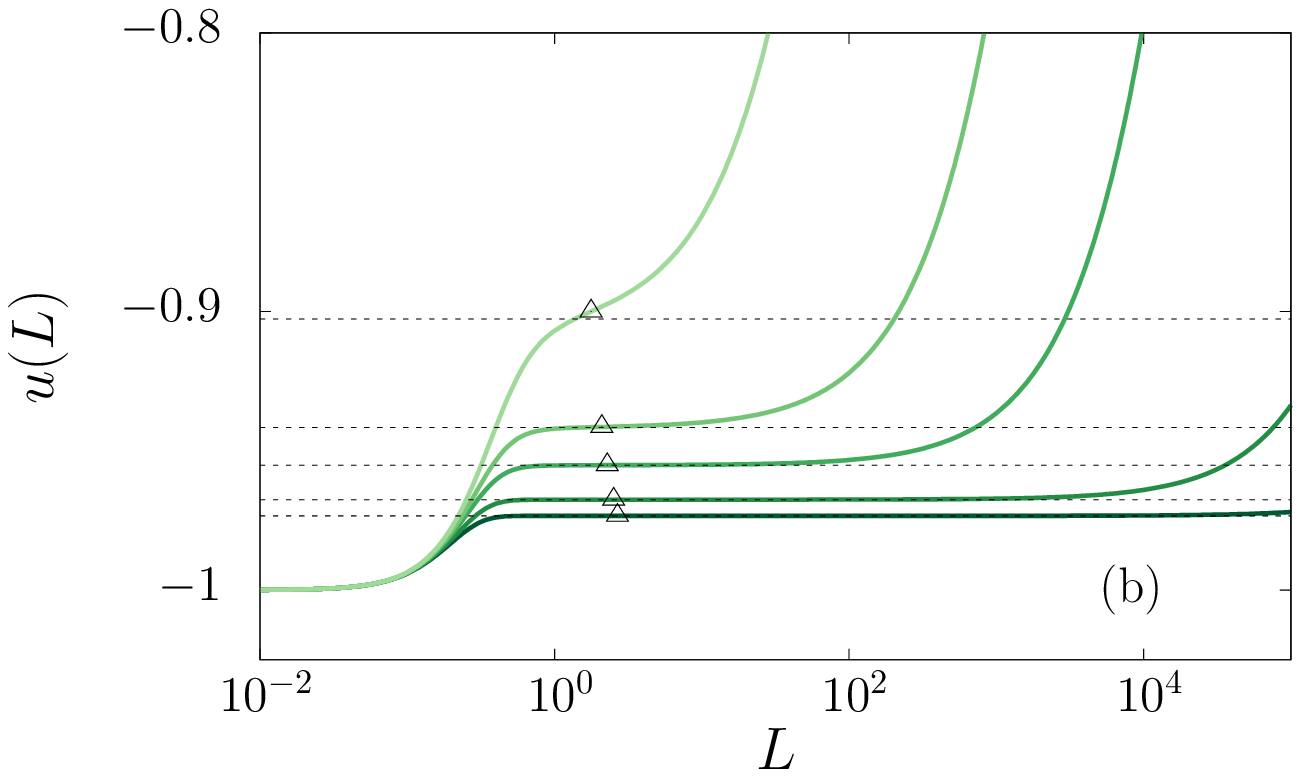} \includegraphics[width=0.47\textwidth]{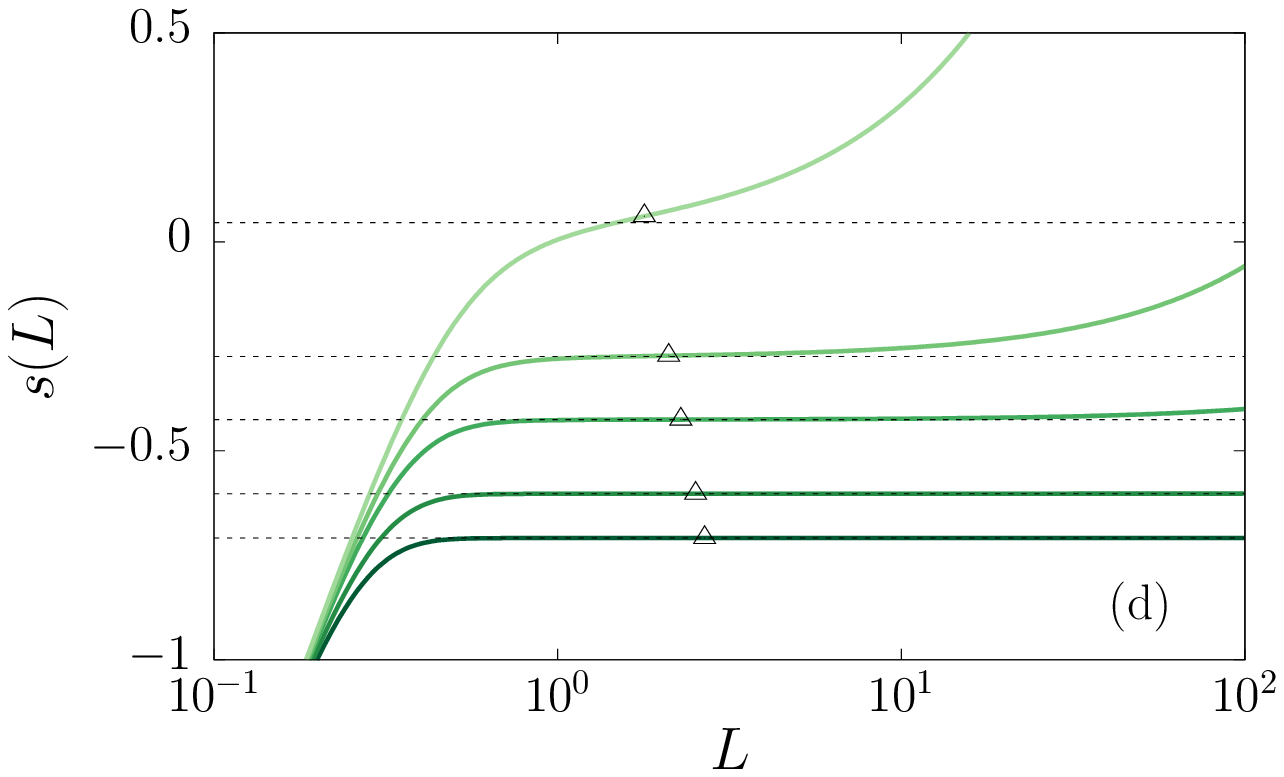}
	\caption{ 
	Dimensionless energy $E$ and entropy $S$ vs. time $t$ (top) and box size $L$ (bottom),  
	in the field $v_4(x)$, so $\mu=4$,
	for different values of $\xi$ indicated in the legend. 
	The time evolution was obtained by integration of the FPE starting from a Dirac delta at the origin, with free boundary conditions, and  $L$-dependent results 
	from the standard normalized  BG state in a box $[-L,L]$.
The dashed horizontal lines show the theoretical prediction of QE values. 
	These are computed with Eq.~(\ref{eq:uQE}) for the energy, and,  
	interestingly,  for the entropy are well described 
	by the equilibrium relation $ {\cal F}/U_0 \equiv f = u - \xi s = \xi \ln Z_0$, 
	where $f$ is the free energy.  
	The symbols indicate the respective  log-inflection points for each observable. 
	}
 	\label{fig:ES}
\end{figure*} 

We have thus regularized the density replacing the diverging partition function with $Z_K$,  
but  the regularization process does not end here. 
As we will soon show, to find the averages of physical observables may require 
additional regularization that depends on the  potential and the observable. 
This is the case of the MSD that we are dealing with.  
Then, before tackling the calculation of its QE value, we start by considering 
any (symmetric) observable  ${\cal O}(x)$ that is integrable with respect
to the non-normalizable Boltzmann factor. In the QE state we have
\begin{equation} \label{eq:O}
  {\cal O}_{\rm QE}  \equiv 
 \langle {\cal O}(x) \rangle_{\rm QE}  = 
\frac{2}{Z_K} \int_0 ^\infty {\cal O}(x) \,{\rm e}^{- v(x)/\xi} dx \,,
\end{equation}
where $Z_K$ is given by Eq.~(\ref{eq:ZR}). 
This formula is very similar to the usual one for calculating equilibrium averages, 
the difference is the regularization of the partition function in the denominator. Let us illustrate this procedure by computing an observable of thermodynamic 
interest such as the average energy $u_{\rm QE}\equiv E_{\rm QE}/U_0 \equiv \langle v(x) \rangle_{\rm QE}$. 
When the potential $v(x)$ decays faster than $1/x$ (hence $K=0$), 
Eq.~(\ref{eq:O}) explicitly becomes
\begin{equation}
u_{\rm QE}   =  
{\int_0^\infty v(x)  {\rm e}^{- v(x)/\xi}    dx \over 
 \int_0 ^\infty \left( {\rm e}^{ - v(x)/\xi} -1 \right) dx  } \equiv  \xi^2 { \partial \ln Z_0 \over \partial \xi}\,. 
\label{eq:uQE}
\end{equation}
It is noteworthy that the QE mean energy obeys the familiar statistical mechanics relation 
with the (regularized) partition function. 
If the potential decays more slowly than $1/x$, like $v_\mu(x)$ with $0<\mu\le 1$, 
the computation of the QE energy needs some modification, 
analogously to Eq.~(\ref{eq:ZR}), namely,  
\begin{equation}
u_{\rm QE}  
=  \frac{2}{Z_K} 
\int_0^\infty v(x) \left(  {\rm e}^{- v(x)/\xi}  
- \sigma_{K-1}(x;\xi) \right)  dx \,, 
\label{eq:ugen}
\end{equation}
 which generalizes Eq.~(\ref{eq:uQE}),  obeying  
$u_{\rm QE} \equiv \xi^2  \partial_\xi \ln Z_K$. 

From the integration of the FPE  (for $v_4(x)$ and 
for different values of $\xi$), 
we show in  Fig.~\ref{fig:ES}
that the stagnation phenomenon occurs for the 
average energy as a 
function of time
\begin{equation}
u(t) \equiv E(t)/U_0    =   \int_{-\infty} ^\infty  v(x) P(x,t) dx,
\end{equation}
as well as for another basic thermodynamic quantity, the dynamical entropy,  
\begin{equation}
s(t) \equiv S(t)/k_B    =   -\int_{-\infty} ^\infty  \ln \big( P(x,t) \big) P(x,t) dx.
\end{equation}
The comparison between theory and numerics for QE states 
is also depicted in Fig.~\ref{fig:ES}, 
showing excellent agreement between time-dependent simulations and 
the QE statistical physics proposed here. 
Interestingly, the QE entropy is computed exactly like in ordinary equilibrium statistical mechanics
$s_{\rm QE} = u_{\rm QE}/\xi  - \ln Z_0$, 
where $Z_0$ is the regularized partition function. 
This shows how regularized statistical mechanics  goes beyond a computation of an effective normalization constant, as thermodynamic identities  are maintained in QE.

\begin{figure}[h!]
\centering
\includegraphics[width=0.5\textwidth]{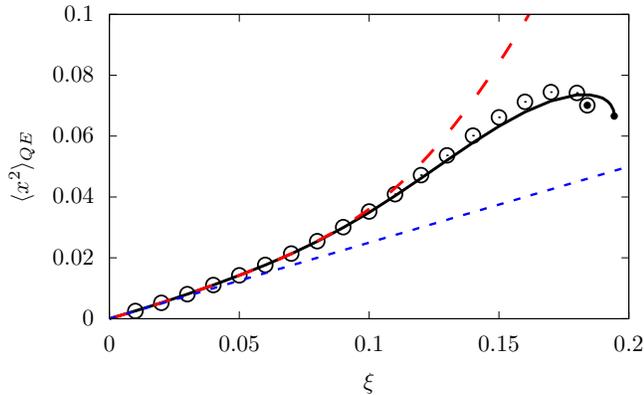}
\caption{ 
Quasi-equilibrium value of the MSD $\langle x^2 \rangle_{\rm QE}$ as a function of the scaled temperature $\xi$. 
We present two theoretical predictions, the first based on Eq.~(\ref{eq:msd}), plotted by a red long-dashed line, and the second one corresponds to the log inflection point $L^\star$ extracted from Eq.~(\ref{eq:msd_L-star}), plotted by the black solid line. 
This latter has a terminus point (small black circle), and beyond this critical value, $\xi \simeq 0.194$,  the  inflection point ceases to  exist.  
It is noteworthy that the log-inflection point in time solutions also presents a critical value $\xi \simeq 0.184$, although 
different from the first one, 
also highlighted by small black filled circle. 
Black hollow circles correspond to the value at the log-inflection point of $\langle x^2(t) \rangle$, obtained from  the FPE   temporal solution.   
 The blue short-dashed line represents the harmonic approximation, drawn for comparison.
 }
\label{fig:MSDxi}
\end{figure}

The moment observables  
$\langle {\cal O}(x)\rangle = \langle x^n\rangle$ require a more careful treatment. 
The integral over $x^n {\rm e}^{  - v(x)/\xi}$ from zero to infinity diverges, so, 
we need to perform a calculation similar to that of Eq.~(\ref{eq:ZR}), 
obtaining
\begin{equation} 
 \langle x^n \rangle_{\rm QE}  = \frac{2}{Z_K} \int_0 ^\infty  
x^n \Bigl( {\rm e}^{-v(x)/\xi} - \sigma_{K'}(x;\xi)  \Bigr)
dx \,,
\label{eq:xn}
\end{equation}
where $K'=\lfloor (n+1)/\mu \rfloor$ and the function $\sigma_{K'}$ regularizes the numerator.  
In the particular case of the MSD presented in Fig.~\ref{fig:msd}, 
$n=2$, and so, for $\mu=4$,  $K'=K=0$. 
Then, Eq.~(\ref{eq:xn}), together with Eq.~(\ref{eq:potential}),  yields the explicit form (see Appendix)
\begin{equation} \label{eq:msd}  
\langle x^2 \rangle_{\rm QE} =      
\frac{
  \  _{2}F_{2} \left( {1 \over 4} , {3 \over 4}; {3 \over 2} , 2 ; {1 \over \xi} \right)  
}{  \  _{2}F_{2} \left( {3 \over 4} , {5 \over 4}; {3 \over 2} , 2 ; {1 \over \xi} \right)   
}\,,
\end{equation} 
plotted by horizontal lines in Fig.~\ref{fig:msd}a, 
in good agreement with numerical results when 
$\xi \lesssim 0.1$.
In this regime, we note that  this theory greatly improves a harmonic approximation of the minimum of the potential, as can be observed in Fig.~\ref{fig:MSDxi}, therefore the nonlinearity of the force field cannot be neglected. 
However, when $\xi \gtrsim 0.1$ we see deviations. 
This is hardly surprising as the stagnation is short-lived, 
moreover, the mentioned condition $\sqrt{t} < {\rm e}^{1/\xi}$ is starting to become questionable. 
 Still  in the next section  we seek  a new method  to describe the intermediate temperature regime.

{\bf Remark:} With numerical simulations  of FPE, we cannot prove that 
log-inflection points vanish beyond a critical value. 
What we see however is that they are impossible to measure precisely, when temperature is $\xi\simeq 0.1$ or higher. In experiments, it would be impossible to detect such log-inflection points at high temperature.    In the next section we will show that a bounded-box method predicts the critical value. This critical temperature separates the low and high temperature limits.

To summarize, the statistical theory we have presented allows the calculation of time-independent averages in QE, 
valid below the critical value of $\xi$. 
The key is that instead of integration over the BG measure, we perform the ensemble average 
over regularized states of the observable. 
The regularization of $Z_K$, which is similar to the standard partition function, 
is given by Eq.~(\ref{eq:ZR}), depending only on the behavior of the potential 
at long distances, while the regularization of the numerator 
in Eqs.~(\ref{eq:uQE})-(\ref{eq:xn}) depends
on the asymptotics of both the potential field and the observable.

\section{Bounded domain approach}
\label{sec:bounded}

So far we have addressed the QE of a time-dependent process in an unbounded domain. 
Can the tools developed so far describe an even wider set of problems, in particular systems of finite size? 
In order to address this issue,  we now consider systems confined in a box of size $L$, 
larger than the trap size, which attain a BG state.  
We will show that for the deep asymptotically flat potentials we are considering herein, the equilibrium state is quasi-independent of $L$, for $1 \ll L \ll   {\rm e}^{1/\xi}$. 
Moreover, the equilibrium state is actually the same as the regularized QE of time-dependent systems with free boundary conditions treated so far, 
for small $\xi$ (see Fig.~\ref{fig:MSDxi}). 
This allows us to add a tool for the calculation of thermodynamic observables in QE, 
and, more importantly, to show the generality of the developed concepts.

We have seen that there is a cutoff time up to which the 
observables of interest are practically time-independent. 
This hints that we may alternatively treat the problem as a time-independent one. 
A step in this direction is to confine the motion to a finite domain  $-L<x<L$. 
Let us consider the potential field  $v(x)$  with reflecting  boundaries at $x=\pm L$.
We focus on the BG equilibrium properties of the particle, so here $t \to \infty$ is considered.
We start by computing the partition function
\begin{equation} 
Z(L) = 2  \int_0 ^{L} {\rm e}^ {-v(x) /\xi} dx\,, 
\label{eq:Z_L}
\end{equation} 
which is of course finite for any large but finite $L$, and then, for the MSD, we find
\begin{equation}
\langle x^2 (L) \rangle = 
\frac{2}{Z(L)}     \int_{0} ^L x^2 \,{\rm e}^{- v(x)/\xi} dx \,. 
\label{eq:msd_L}
\end{equation} 
This is plotted as a function of $L$ in Fig. \ref{fig:msd}b, for several values of $\xi$.
Notice, as we will demonstrate below, 
that the stagnation values are the same as those found from the time dependent solution.
This is remarkable as it allows us to obtain the QE semi-analytically without  
evolving the system in time. 
Moreover,  Eq.~(\ref{eq:msd_L}) allows to obtain theoretically the inflection point in the 
stagnation region, 
also plotted in Fig. \ref{fig:msd}b.


First we rewrite the integral for $Z(L)$, by adding and subtracting terms, as
\begin{equation} \nonumber
Z(L) = 2  \int_0 ^{L} \Bigl( {\rm e}^{-v(x)/\xi} - \sigma_K(x;\xi) \Bigr) dx + 2 \int_0 ^{L} \sigma_K(x;\xi) dx \,.
\end{equation}
Then, we split $Z(L)$ as
\begin{equation}
Z(L)= Z_K +Z_K^>(L)+Z_K^<(L) \,,
\end{equation}
where
\begin{eqnarray} \label{Zinf}
Z_K &=&  2\int_0 ^\infty \Bigl( {\rm e}^{-v(x)/\xi} - \sigma_K(x;\xi)  \Bigr) dx \,, \\ \label{Z>}
Z_K^> (L) &=& - 2\int_L ^\infty \Bigl( {\rm e}^{-v(x)/\xi} - \sigma_K(x;\xi)  \Bigr) dx \,,\\ \label{Z<}
Z_K^< (L) &=&   2\int_0 ^{L}  \sigma_K(x;\xi) dx \,.
\end{eqnarray}  
Notice that the convergent integral  $Z_K$ is  $L$-independent.
It is easy to show that $Z_K^>(L)$ is relatively  small when $L\gg 1$. 
Clearly $Z_K^<(L)\sim 2 L$, 
indeed when we fix $\xi$ and take  $L\to \infty$, this is the leading contribution.
However, if we consider $\xi\ll 1$,  with fixed though large  $L$, then 
$ Z_K \propto {\rm e}^{1/\xi}$ dominates. Since it is $L$-independent,  
it is well-suited for the description of an infinite system, where there is stagnation in time, 
i.e., those systems modeled by the FPE in  QE but with free boundary conditions.

We now perform an analogous splitting  of the integral in the numerator of Eq.~(\ref{eq:msd_L}).  
Namely, we introduce the factor $x^2$ in the integrands of Eqs.~(\ref{Zinf})-(\ref{Z<}), substitute $K\to K'$, 
and perform a split of the numerator of Eq.~(\ref{eq:msd_L}), in the respective integrals 
$X_{K'}$, $X_{K'}^>$ and $X_{K'}^<$ 
(note that $X$ in the original units  has dimension  of a cubic length).
In the limit $L\to \infty$ and  fixed $\xi$, we find
the trivial leading term $X_{K'}^< \sim  L^3 /3$, and this together with
the normalization gives $\langle x^2(L) \rangle \propto L^2$ as expected.
Meanwhile $X_{K'}^>$ is small and negligible when $L \gg 1$. 
Again, since we are interested in the stagnation limit, if $L$ is not too large   
the leading term is  
$X_{K'} = 2 \int_0 ^\infty  
x^2 \Bigl( {\rm e}^{-v(x)/\xi} - \sigma_{K'}(x;\xi)  \Bigr)
dx$, which is also proportional to ${\rm e}^{1/\xi}$. 

Putting all these results together, we arrive at   
$\langle x^2\rangle_{\rm QE} = X_{K'}/Z_K$, that is,
\begin{equation} 
 \langle x^2 \rangle_{\rm QE}  = \frac{2}{Z_K} \int_0 ^\infty  
x^2 \Bigl( {\rm e}^{-v(x)/\xi} - \sigma_{K'}(x;\xi)  \Bigr)
dx \,,
\label{eq:x2}
\end{equation}
which is   $L$ independent and hence also $t$ independent. 
Notice that this expression precisely coincides with   Eq.~(\ref{eq:xn}) for $n=2$, which was obtained
through the time-dependent solution in the unbounded domain, 
illustrating how both approaches meet.

\begin{figure}[h!]
\centering
\includegraphics[width=0.5\textwidth]{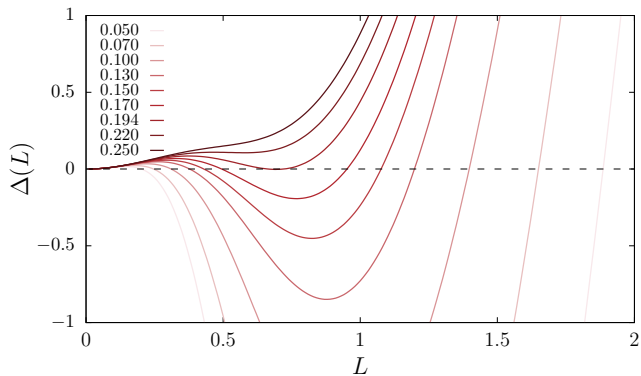}
\caption{ 
Numerically solving Eq.~(\ref{eq:msd_L-star}), $\Delta(L)=0$, for the potential $v_4(x)$ ($\mu=4$) and different values of the scaled temperature $\xi$ in the legend. Notice that for small $\xi$ there are two solutions, besides the trivial solution $L=0$, which collapse at a critical value ($\xi\simeq 0.194$ the case of the figure) and vanish at higher temperature. 
}
\label{fig:delta}
\end{figure}
 
\subsection{Critical temperature} 
 
Still considering the $L$-dependent expression  $\langle x^2 (L) \rangle$ in Eq.~(\ref{eq:msd_L}), we can obtain the log-inflection 
$L^\star$,  through $d^2 \langle x^2(L)\rangle d(\log L)^2=0$, 
obtaining the transcendental equation
\begin{equation}
\label{eq:msd_L-star} 
2 L^2 -  (L^2 - \langle x^2 \rangle) \left\{ \frac{4 L {\rm e}^{-v(L)/\xi}}{Z(L)}  -\frac{Lf(L)}{\xi} - 1 \right\} \equiv \Delta(L)=0.
\end{equation}
We numerically find the solutions of 
Eq.~(\ref{eq:msd_L-star}), as the zeros of 
$\Delta(L)$, which is plotted vs. $L$ in  
Fig.~\ref{fig:delta}  for different values $\xi$. Besides the trivial solution 
($L^\star=0$, which is nonphysical), there are two solutions for small $\xi$, that collide at a 
critical value [$\xi \simeq 0.194$, when $\mu=4$ in Eq.~(\ref{eq:potential})] and cease 
to exist above that temperature.  
Roughly this $\xi$ marks the transition between low and high temeperature regimes, 
and it is observable-dependent and in that sense not universal. 

\begin{figure}[h!]
\centering
\includegraphics[width=0.5\textwidth]{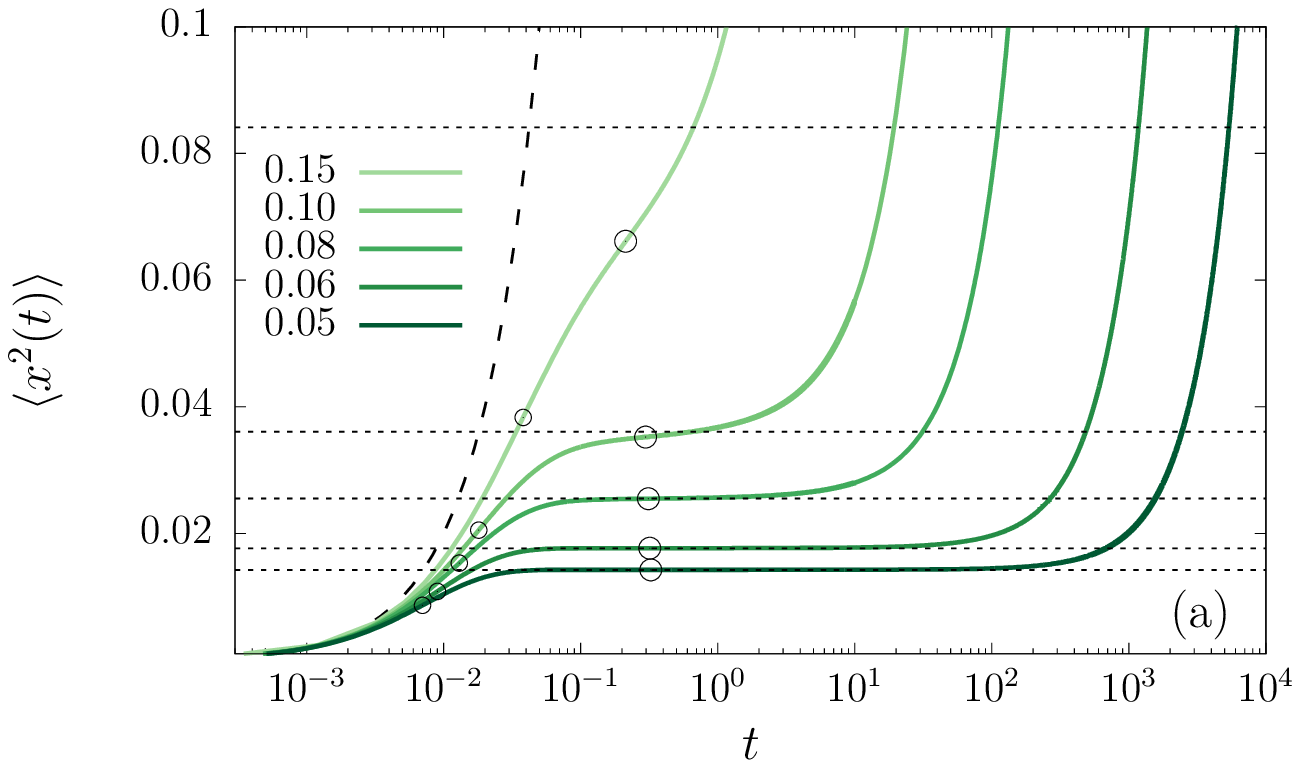}
\includegraphics[width=0.5\textwidth]{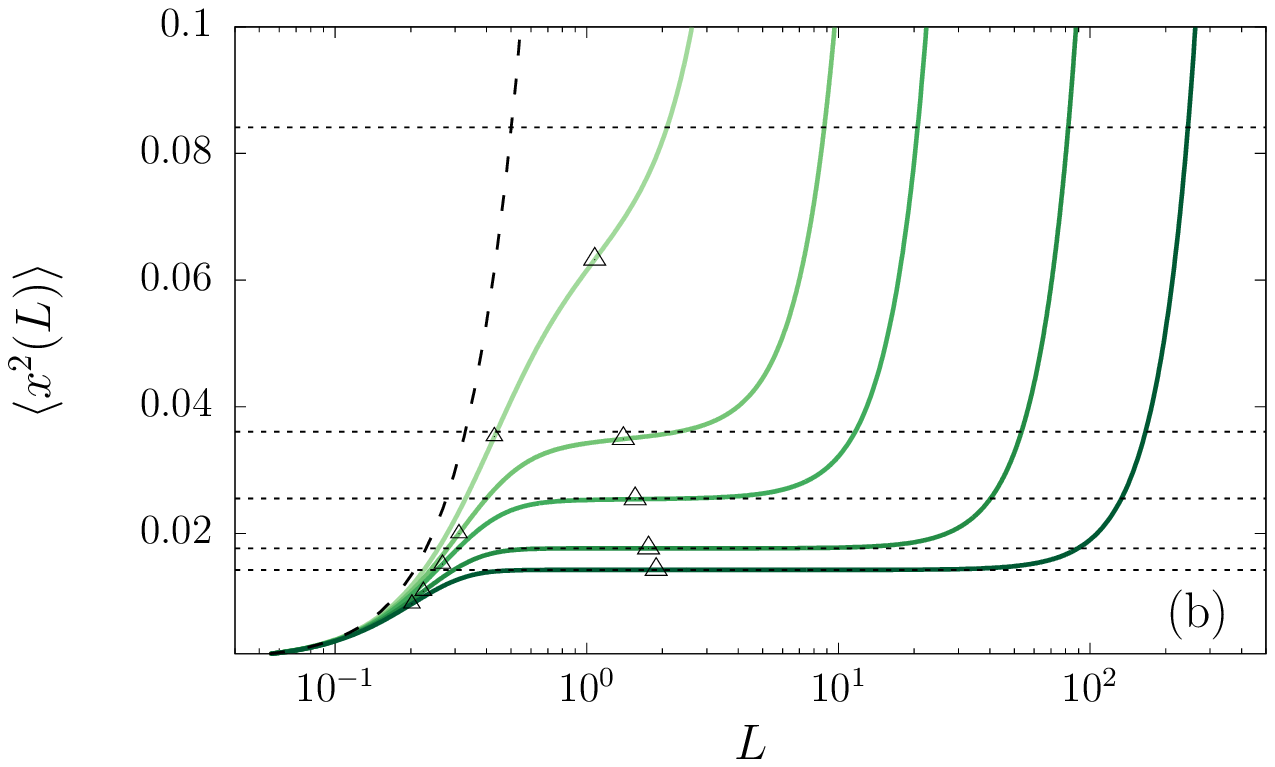}
\caption{ 
MSD   vs. time $t$ (a) and vs. $L$ (b). The same data of 
Fig.~\ref{fig:msd} are plotted in linear vs. logarithmic scale, to display the changes of concavity associated to the inflection points. 
In (b), the nontrivial inflection points, solution 
of Eq.~(\ref{eq:msd_L-star}) are highlighted. 
At low temperature, an  
inflection point (small symbol) predicts the 
start of stagnation, while another
(large symbol) lies in the stagnation zone. 
When the temperature  $\xi$ increases, 
these two  inflection points meet and disappear above a critical temperature. The black dashed line represents the free-particle case 
(limit $\xi \to \infty$), which has no inflection points. 
It is interesting that similar features are observed 
in the time evolution, shown in panel (a). 
}
\label{fig:MSDlinear}
\end{figure}

In Fig.~\ref{fig:MSDlinear}, 
we plot the MSD versus time $t$ and box size $L$ on a 
linear-log scale (unlike Fig.~\ref{fig:msd}, which is a 
log-log plot).  In this figure we present the two inflection  points, which as mentioned are found only for low enough $\xi$. 
At low temperatures, there are two nontrivial inflection points. The first anticipates the start of the stagnation, 
while the second, that we call $L^\star$, lies in the stagnation region. 
We take the latter as a signature of QE state 
(see Fig.~\ref{fig:msd}). 
Above a certain temperature, both inflection points merge and disappear in Fig.~\ref{fig:MSDlinear}. The line corresponding to the infinite temperature limit (no potential), where clearly there are no inflection points at all is also plotted for comparison.

\subsection{Relation between the two approaches} 

When $\xi$ is small enough, we observe that (see Fig. \ref{fig:MSDxi})
$\langle x^2 \rangle_{\rm QE}  \simeq \langle x^2  (L^\star) \rangle \simeq   \langle x^2  (t^\star) \rangle$,
where the `*' stands for the (largest) log-inflection point, 
which characterizes the stagnation zone 
(see Fig.~\ref{fig:msd}). 
Also the mean energy has two nontrivial inflection points, the second one  in the stagnation region (shown in Fig.~\ref{fig:ES}), while
in the case of the entropy we have a single inflection point in the stagnation region.

Thus, more generally, we postulate that for an observable ${\cal O}$, we have, 
when QE exists, 
\begin{equation}
\langle {\cal O} \rangle_{\rm QE}  
\simeq \langle {\cal O} (L^*) \rangle \simeq   \langle {\cal O} (t^*) \rangle\ \,.
\end{equation} 
Thus, we can use the finite-size system expression for the observable, 
obtain $L^*$ and then infer the temporal stagnation value. 
For example: in Fig.~\ref{fig:MSDlinear}, we use Eq.~(\ref{eq:msd_L-star}) for the MSD to 
determine $L^*$. 
We then compare the result with the time-dependent calculation of the MSD,
using the FPE, showing good agreement between them, for small $\xi$. 
As mentioned, with this method, there is no need to evolve the system in time, 
which is very important since the escape time is exponentially large.

Figure~\ref{fig:MSDxi} also reveals that, 
for small $\xi$, there is excellent agreement between the box method and the theory of QE. 
This is the regime where we have a clear separation of time scales in the dynamics. 
When $\xi\gtrsim 0.1$ we observe deviation between the inflection-point and the QE approaches.
We see from the figure that the time-dependent simulation results are  very close to
those obtained by the bounded domain approach  Eq.~(\ref{eq:msd_L}) even when $\xi$ is relatively  large.

\section{Final remarks}
\label{sec:final}

We have shown that long-lived QE states emerge when particles are subject to  
an external field which has a deep well at the origin and is asymptotically flat.
Despite the divergent character of the standard partition function  
due to the non-confining potential, a regularization 
procedure is still possible, allowing one to calculate quantities 
in the QE states along the lines of the standard 
recipes of statistical mechanics. 
The  regularization strategy can be applied to a vast number of observables, 
in particular thermodynamic quantities, e.g., the energy, entropy and free energy.
Experimental  proof of these concepts can be made for a single molecule weakly attached to a membrane or a surface embedded in a thermal bath~\cite{krapf19,krapf16,wang17}.

As we increase temperature, stagnation becomes short lived. 
We identified a critical temperature that marks the transition between low and high temperature states. 
At that temperature, two merging inflection points  collide.  
In this sense, the inflection point method  not only defines the QE low temperature states. 
The concept is very useful also far from this limit. 
Indeed when departing from the low temperature regime, we noticed that inflection points 
based on finite systems are the useful theoretical tool to describe the QE phenomenon. 

In general,  metastable states emerge in systems with clear separation of time scales, 
and then observables appear stationary. Eventually a true equilibrium state 
can be reached~\cite{kurchan,Garrahan}, for example if the system is finite. 
While in our case there is a transition from a metastable state to a diffusive behavior, 
the tools developed here will turn out useful also in other situations. 
For example, if we have confinement of the particles in a very  large box, 
the initial QE will not be influenced by finite-size effects, 
as long as the observation time scale  is shorter than the time of escape from the well. 
The investigation of the relaxation properties, the underdamped dynamics,  systems with several metastable states,  anomalous  processes satisfying the fluctuation dissipation theorem, 
and interacting systems is left for future work.

{\bf Acknowledgments:} D.K. and E.B. acknowledge 
the support of Israel Science Foundation's grant 1898/17. 
L.D. and C.A. acknowledges partial financial support  
from  Conselho Nacional de Desenvolvimento Cient\'{\i}fico
e Tecnol\'ogico (CNPq), Funda\c c\~ao de Amparo \`a Pesquisa do Estado do Rio de Janeiro (FAPERJ), and Coordena\c c\~ ao de Aperfei\c oamento de Pessoal de N\' ivel Superior - Brasil (CAPES) - Finance Code 001.


\appendix

\section{Appendix}
\label{sec:appendix}

Let us consider the particular case of 
Eq.~(\ref{eq:xn}) in which $n=2$, corresponding to the MSD,  
and $\mu=4, (K^\prime=K=0)$ corresponding to the potential field $v_4(x)$. 
From Eq.~(\ref{eq:sigma}), using that $\sigma_0(x;\xi)=1$,  explicitly we have
\begin{eqnarray} \nonumber
 \langle x^2 \rangle_{\rm QE}  
 &=& \frac{2}{Z_0} \int_0 ^\infty  
 \Bigl( {\rm e}^{-v_4(x)/\xi} - 1 \Bigr)
dx   \\
&=&\frac{ \int_0 ^\infty  
x^2 \Bigl( {\rm e}^{-v(x)/\xi} - 1 \Bigr)
dx }{\int_0 ^\infty  
 \Bigl( {\rm e}^{-v(x)/\xi} - 1 \Bigr)
dx }\,.
\label{eq:x2A}
\end{eqnarray}
where  $v_4(x)= -1/(1+x^2)^2$.
We Taylor expand  the exponential in the integral in powers of $v_4(x)/\xi$ and integrating term by term. 

For the partition function $Z_0$, using that
\begin{eqnarray} \nonumber
    \int_0^\infty [-v_4(x)/\xi]^k  dx &=&
    \int_0^\infty  [\xi(1+x^2)^2]^{-k} \, dx  \\
    &=&
    \frac{1}{\xi^k}\frac{\sqrt{\pi}\,\Gamma[2k-1/2]}{2 \, \Gamma[2\,k]},
\end{eqnarray}
and the $\Gamma$ function, we obtain

\begin{eqnarray} \nonumber
Z_0 &=&  2\int_0 ^\infty  \Bigl( {\rm e}^{-v_4(x)/\xi} - 1 \Bigr)
dx    \\  \nonumber
&=& 2\sum_{k=1}^\infty  \frac{1}{k!} \int_0^\infty [-v_4(x)/\xi]^k dx =
   \sum_{k=1}^\infty  \frac{1}{k!}  \frac{1}{\xi^k}\frac{\sqrt{\pi}\,\Gamma[2k-1/2]}{ \Gamma[2k]}   \\ \nonumber
&=&  \frac{\pi}{2 \xi} \,{}_{2}F_{2} \left( {3 \over 4} , {5 \over 4}; {3 \over 2} , 2 ; {1 \over \xi} \right) \,.
 \end{eqnarray}

 For the denominator, we apply the same procedure, using
  
\begin{eqnarray} \nonumber
    \int_0^\infty x^2 [-v_4(x)/\xi]^k  dx &=&
    \int_0^\infty  x^2[\xi(1+x^2)^2]^{-k} \, dx \\
    &=&
    \frac{1}{\xi^k}\frac{\sqrt{\pi}\,\Gamma[2k-3/2]}{4\,\Gamma[2k]},
\end{eqnarray}
then
\begin{eqnarray} \nonumber
&&2\int_0 ^\infty  \Bigl( x^2({\rm e}^{-v_4(x)/\xi} - 1) \Bigr)
dx =  2\sum_{k=1}^\infty  \frac{1}{k!} \int_0^\infty x^2 [-v_4(x)/\xi]^k dx \\ \nonumber
&=&
   \sum_{k=1}^\infty  \frac{1}{k!}  \frac{1}{\xi^k}\frac{\sqrt{\pi}\,\Gamma[2k-3/2]}{ 2\Gamma[2k]} 
  = \frac{\pi}{2 \xi} \,{}_{2}F_{2} \left( {1 \over 4} , {3 \over 4}; {3 \over 2} , 2 ; {1 \over \xi} \right) \,.
 \end{eqnarray}
Combining the expression for the numerator and denominator of Eq.~(\ref{eq:x2A}), we arrive to Eq.~(\ref{eq:msd}).

\end{document}